\documentclass[aps,prx,amsmath,amssymb,superscriptaddress,notitlepage,groupedaddress,twocolumn,reprint]{revtex4-1}
\usepackage{chngcntr}\usepackage{graphicx}
\usepackage{dcolumn}
\usepackage{stackengine}
\usepackage[section]{placeins}
\usepackage[dvipsnames*,svgnames]{xcolor}
\usepackage{verbatim}
\usepackage[position=bottom,caption=false,captionskip=3pt,font=normalsize,subrefformat=simple,labelformat=simple]{subfig}

\usepackage{listings}
\usepackage{mathtools}
\usepackage{todonotes}
\usepackage{amsmath}
\usepackage{chngcntr}
\usepackage{bm}
\usepackage{url}

\usepackage[breaklinks]{hyperref}
\usepackage{breakurl}
\usepackage{tabularx}
\usepackage{soul}
\usepackage{braket}
\hypersetup{
colorlinks,
linkcolor={blue!100!black},
citecolor={blue},
urlcolor={blue!80!black}}
\def\equationautorefname~#1\null{#1\null}
\usepackage[bottom]{footmisc} \makeatletter \def\p@section{} \def\p@subsubsection{} \makeatother
\begin{document}
\title{Phonon Transport in Patterned Two-Dimensional Materials from First Principles}

\author{Giuseppe Romano}
\email{romanog@mit.edu}
\affiliation{Department of Mechanical Engineering, Massachusetts Institute of Technology, 77 Massachusetts Avenue,
Cambridge, Massachusetts 02139, USA}

\begin{abstract}

Phonon size effects induce ballistic transport in nanomaterials, challenging Fourier's law. Nondiffusive heat transport is captured by the Peierls-Boltzmann transport equation (BTE), commonly solved under the relaxation time approximation (RTA), which assumes diagonal scattering operator. Although the RTA is accurate for many relevant materials over a wide range of temperatures, such as silicon, it underpredicts thermal transport in most two-dimensional (2D) systems, notoriously graphene. Here we present a formalism, based on the BTE with the full collision matrix, for computing the effective thermal conductivity of arbitrarily patterned 2D materials. We apply our approach to porous graphene and find strong heat transport suppression in configurations with feature sizes of the order of micrometers; this result, which is rooted in the large generalized phonon MFPs in graphene, corroborates the possibility of strong thermal transport tunability by relatively coarse patterning. Lastly, we present a promising material configuration with low thermal conductivity. Our method enables the parameter-free design of 2D materials for thermoelectric and thermal routing applications. 

\end{abstract}
\maketitle
\section{INTRODUCTION}
Phonon boundary scattering in nanostructures facilitates control over heat flow, enabling several thermal-related applications, including thermoelectrics~\cite{Vineis,marconnet2013casimir}, heat guiding and lensing~\cite{anufriev2017heat}. Modeling accurately these systems, however, is particularly challenging because of the nondiffusive nature of heat transport~\cite{vega2016thermal}; in fact, when the characteristic length becomes comparable with the phonon mean free paths (MFPs), ballistic transport may dominate~\cite{Ziman2001,chenbook} and the Boltzmann transport equation (BTE) must be used~\cite{Majumdar1993am}. A common semplification to the BTE is given by the relaxation time approximation (RTA) where phonon repopulation is neglected (the scattering operator is assumed diagonal). While the RTA has been proven reliable in relevant systems, such as Si~\cite{esfarjani2011heat}, it severely underpredicts the thermal conductivity of several two-dimensional materials, including graphene~\cite{fugallo2014thermal}, boron nitride~\cite{lindsay2011enhanced} and molybdenum disulphide~\cite{cepellotti2015phonon}. In fact, in these systems the scattering operator has significant off-diagonal components and size effects are dictated by the ``generalized'' phonon MFPs, significantly larger than the standard MFPs~\cite{li2014shengbte}. Lastly, the RTA cannot capture phonon hydrodynamic effects and second sound~\cite{huberman2019observation}.

Several models have been developed to compute space-resolved thermal transport beyond the RTA. Cepellotti et al~\cite{cepellotti2015phonon} have applied the concept of relaxons to model heat transport in molybdenum disulfide nanoribbons~\cite{Cepellotti2017BoltzmannEffect}. Landon et al ~\cite{landon2014deviational} and Mei et al \cite{Mei2014Full-dispersionNanoribbons}, solved the space-dependent BTE with the phonon Monte Carlo obtaining good agreements with experiments on graphene nanoribbons (GNRs). Marepalli et al ~\cite{marepalli2019spectrally} solved the BTE in GNRs with short widths, using Tersoff interatomic potentials, and taking into account phonon confinement. In ~\cite{luo2019direct}, Luo et al, computed second-sound in GNRs by solving the space-dependent Callaway model~\cite{callaway1959model}. An analytical approach based on Green's functions has been reported in~\cite{chiloyan2017micro}, where the non-homogeneous BTE is solved also including arbitrary heat sources. Torres et al~\cite{Torres2017FirstSemiconductors} developed the ``kinetic-collective'' model, which is based on the separate contributions from Normal and Umklapp scattering. More recently, Varnavides et al~\cite{varnavides2019nonequilibrium} developed the iterative, space-dependent BTE and applied it to silicon nanoparticles and Si/Ge interfaces, with the perturbation being a constant heat source.

We present a framework to compute the effective thermal conductivity of two-dimensional materials with complex shapes from nano- to macro- scales. Our method combines the space-dependent BTE~\cite{varnavides2019nonequilibrium} and the concept of effective phonon temperature~\cite{romano2015} to calculate the spatially-resolved thermal flux upon the application of a difference of temperature. Furthermore, we identify the solution to the standard diffusive equation as the first guess to our solver; this choice crucially speeds up convergence, enabling simulations that would be prohibitive otherwise. Special emphasis is given to the energy conservation of the energy operator. We apply our method to the prototypical 2D material graphene with different geometries, including nanoribbons and porous configurations. Due to the generalized phonon MFPs, being tents of micrometers at room temperature, we obtain significant reduction in the thermal conductivity even for patterning with a resolution as coarse as 1 $\mu$m. Lastly, we identify a material with very low thermal conductivity, hence appealing to thermoelectric applications. Since the scattering operator includes also momentum-conserving events, our method naturally includes hydrodynamic transport~\cite{cepellotti2015phonon,lee2015hydrodynamic}. 

\section{Boltzmann transport equation}
Steady-state phonon transport is described by the linearized Boltzmann transport equation (BTE)~\cite{Ziman2001} 
\begin{equation}
\begin{split}\label{bte}
-\mathbf{v}_{\mu}\cdot \nabla n_\mu (\mathbf{r})= \sum_{\nu} \Omega_{\mu \nu} \left[n_\nu(\mathbf{r}) - \bar{n}_\nu(\mathbf{r}) \right],
\end{split}
\end{equation}
where $n_\mu(\mathbf{r})$ is the space-dependent non-equilibrium phonon distribution, $\mathbf{v}_\mu$ is the group velocity, $\Omega_{\mu\nu}$ is the collision matrix, and $ \bar{n}_\mu (T)$ is the Bose-Einstein distribution at temperature $T(\mathbf{r})$, given by $\bar{n}_\mu (T) = \left[ e^{\frac{\hbar\omega_\mu}{k_B T(\mathbf{r})}}-1\right]^{-1}$, with $\omega_\mu$ being the phonon angular frequency. The mode label $\mu$ collectively describes wave vector $\mathbf{q}$ (running up to $N$) and polarization $p$. Equation~\ref{bte} balances out the change in phonon population due to drift (left hand side) and collisions (right hand side). In passing, we note that since the BTE does not take into account resonances or coherent effects, all the results presented below can be considered object for experimental verification only for feature sizes larger than 100 nm~\cite{Cahill2003NanoscaleTransport,Munoz2010BallisticRibbons}. Nevertheless, for the sake of completeness we provide data also for smaller dimensions. Furthermore, phonon tunnelling across branches~\cite{Simoncelli2019UnifiedGlasses} is neglected and beyond the scope of this work. 

Eq.~\ref{bte} can be simplified if we assume small variation of $T$ with respect to the ambient temperature $T_0$, so that the equilibrium term can be Taylor expanded to its first order in $T(\mathbf{r})-T_0$, i.e. $\bar{n}_\mu(T) = \bar{n}_\mu(T_0) + C_\mu \left(\hbar \omega_\mu\right)^{-1}\left(T(\mathbf{r})-T_0\right)$; the term $C_\mu$ is the mode-resolved heat capacity, given by $C_\mu = k_{\mathrm{B}} \eta_\mu^2 \left(\sinh{\eta_\mu}\right)^{-2}$, where $\eta_\mu = \hbar \omega_\mu /\left(2 k_\mathrm{B}T_0\right)$. 

In the following, for clarity we drop out any space dependence. Conveniently, we define the effective phonon temperature as $T_\mu = T_0 + \hbar \omega_\mu C_\mu^{-1} \left[n_\mu - \bar{n}_\mu (T_0)  \right]$. The BTE then becomes  
\begin{equation}
\begin{split}\label{new_bte}
-\mathbf{S}_\mu \cdot\nabla T_\mu = \sum_\nu W_{\mu \nu} \left[ T_\nu - T\right],
\end{split}
\end{equation}
where $ W_{\mu\nu} = \Omega_{\mu\nu} \omega_\mu C_\nu \omega_\nu^{-1}V^{-1}N^{-1} $ and $S_\mu^\alpha = C_\mu v_\mu^{\alpha}V^{-1}N^{-1}$. In deriving Eq.~\ref{new_bte}, we have used the normalization factor $V_n N$, where $V_n$ is the $n$-dimensional volume of the unit-cell of the crystal lattice. We note that the matrix $W_{\mu\nu}$ is symmetric. In fact, the collision operator in Eq.~\ref{bte} can be rewritten as $\Omega_{\mu\nu} = A_{\mu\nu} /\bar{n}_\nu(T_0)/(\bar{n}_\nu(T_0) + 1 )$, where $A_{\mu\nu}$ is a symmetric positive semidefinite matrix~\cite{Fugallo2013AbConductivity}; using the expression for $C_\nu$, we have $W_{\mu \nu} = \left(V_n N k_B T_0^2/\hbar^2\right)^{-1} A_{\mu \nu}\omega_\mu \omega_\nu$. The matrix $A_{\mu\nu}$ is energy conserving, i.e. $\sum_\mu \omega_\mu A_{\mu\nu} = 0$~\cite{Landon2014ARedacted}. Furthermore, from $A_{\mu\nu}=A_{\nu\mu}$ it follows that $\sum_\nu \omega_\nu A_{\mu\nu} = 0$; consequently, the sum of each row and column of $W_{\mu\nu}$ is zero. However, in practice, energy conservation is not strictly conserved due to the approximation of the Dirac functions in $A_{\mu\nu}$ ~\cite{landon2014deviational} (see Appendix A). As reported in Appendix B, we enforce strict energy conservation using the Lagrange multipliers method~\cite{Landon2014ARedacted}. In light of this discussion, the right hand side of Eq.~\ref{new_bte} simply becomes $\sum_{\nu} W_{\mu\nu} T_\nu$. In passing, we note that the left hand side of Eq.~\ref{new_bte} can be recast into $\nabla\cdot\mathbf{J}_\mu$, where $\mathbf{J}_\mu = T_\mu \mathbf{S}_\mu$ is the mode-resolved thermal flux, normalized by $V_n N$. Then, summing both sides over the index $\mu$ we obtain vanishing divergence of the total heat flux, another instance of energy conservation. Conversely, within the relaxation time approximation, $W_{\mu\nu} = \delta_{\mu\nu}C_\nu/\tau_\nu$, and Eq.~\ref{new_bte} is not energy conserving. A common remedy is to define $T$ such that $\sum_\mu C_\mu/\tau_\mu\left(T_\mu-T\right) =0 $, giving $T = \left[ \sum_\mu C_\mu/\tau_\mu   \right]^{-1} \sum_\mu \left(C_\mu/\tau_\mu\right) T_\mu$~\cite{chiloyan2017micro,romano2015,Hua2014AnalyticalEquation,Carrete2017AlmaBTEMaterials,Zhang2019AnPolarization}.





\begin{figure*}
\subfloat[\label{fig0a}]
{\includegraphics[width=0.48\textwidth]{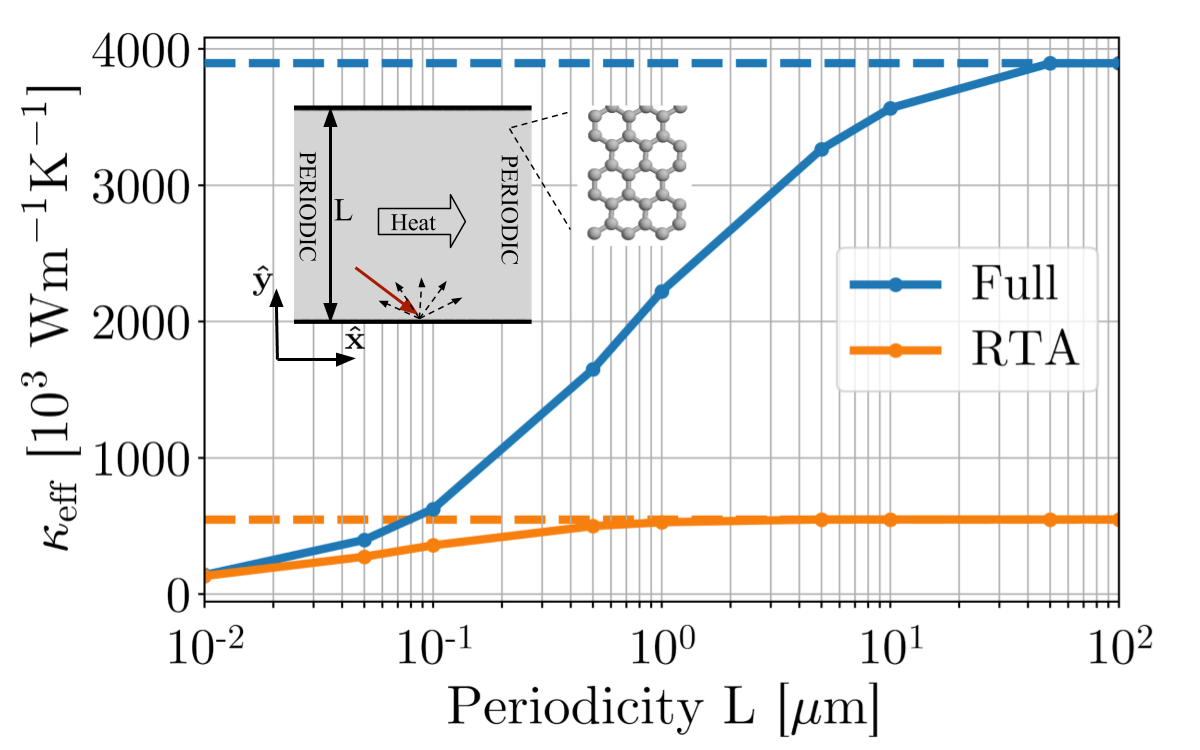}}
\hfill
\subfloat[\label{fig0b}]
{\includegraphics[width=0.48\textwidth]{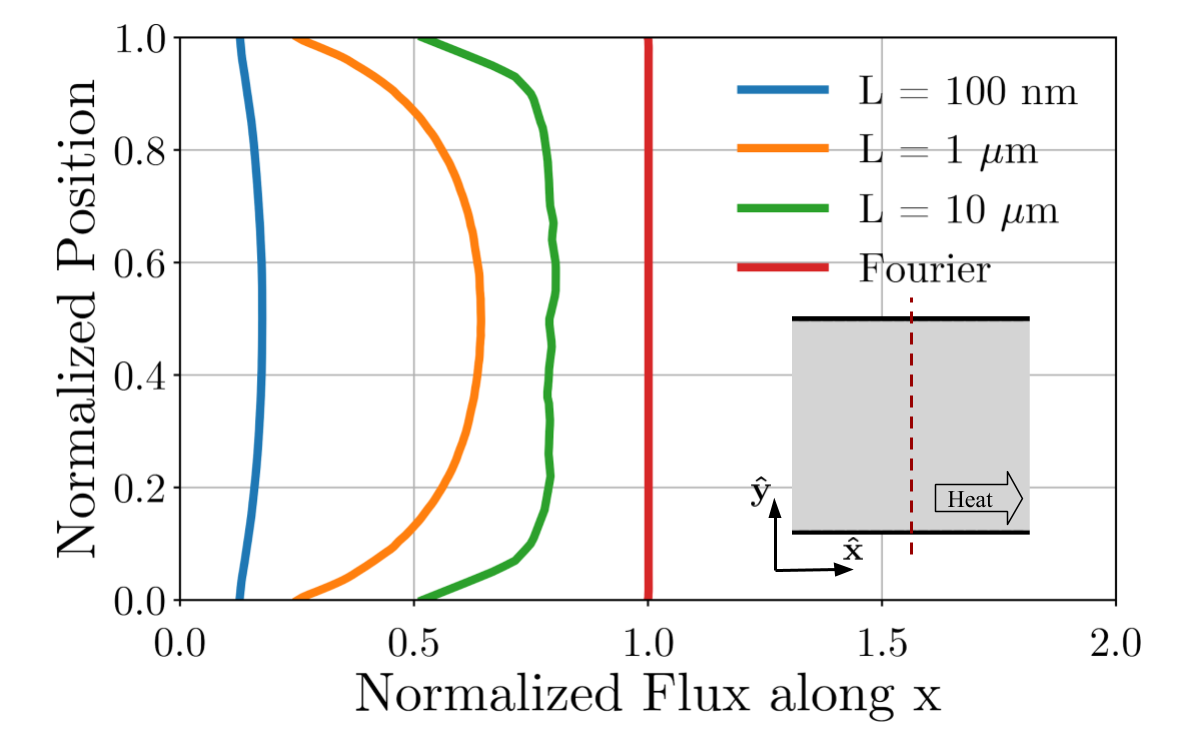}}
\caption{a) Effective thermal conductivity of graphene nanoribbons with infinite lengths and varying width $L$, for both RTA and Full cases. The dashed lines correspond to bulk values. Thermal flux is enforced along the \textit{x}-axis. b) A cut to the $x$-component of thermal flux along the direction perpendicular to the applied temperature gradient.}
\end{figure*}

To solve Eq.~\ref{new_bte}, we decompose the collision matrix into two terms, $W_{\mu\nu} = (V_n N)^{-1} C_\mu /\tau_\mu \delta_{\mu\nu} + W^{\mathrm{od}}_{\mu\nu}$ \cite{Ziman2001,Li2012ThermalPrinciples,Fugallo2013AbConductivity,Varnavides2019NonequilibriumInterfaces}, where $W^{\mathrm{od}}_{\mu\nu}$ is the off-diagonal term. The BTE, then, becomes  
\begin{equation}
\begin{split}\label{final}
\mathbf{F}_\mu \cdot \nabla T_\mu^{(n)} + T_\mu^{(n)} = -\frac{\tau_\mu}{C_\mu}\sum_\nu W^{\mathrm{od}}_{\mu\nu} T_\nu^{(n-1)},
\end{split}
\end{equation}
where $\mathbf{F}_\mu = \mathbf{v}_\mu \tau_\mu$; similarly to~\cite{Omini1995AnConductivity,li2014shengbte,Varnavides2019NonequilibriumInterfaces}, Eq.~\ref{final} is solved iteratively. The temperature formulation of the BTE facilitates the connection with Fourier's law, employed here to compute a first guess to the mode temperatures; that is, $T_\mu^{(0)}$ is the result of the standard diffusive heat conduction equation $ \nabla^2 T_F = 0$. Being mode-independent, the right hand side of Eq.~\ref{final} at the zeroeth iteration becomes $-T_F \tau_\mu/C_\mu \sum_{\mu\nu}W^{\mathrm{od}}_{\mu\nu}=T_F$. The equation for $T_\mu^{(1)}$ then becomes $\mathbf{F}_\mu\cdot\nabla T_\mu^{(1)} + T_\mu^{(1)} = T_F$, i.e. a reminiscence of the RTA~\cite{romano2015}. The choice of $T_F$ as a proxy for $T_\mu^{(0)}$ and the temperature formulation of the BTE in Eq.~\ref{final} are the first results of this paper. In passing, we note that for large structures, Eq.~\ref{final} is not guaranteed to converge and a relaxation factor (in our work we use 0.75) must be used~\cite{Varnavides2019NonequilibriumInterfaces}; This issue is due the fact that the eignvalues of $W_{\mu\nu}$ are not guaranteed to be smaller than 1~\cite{cepellotti2016thermal}.

Equation~\ref{final} is performed by the upwind finite volume method~\cite{romano2011multiscale,murthy2005review}, implemented within the OpenBTE framework~\cite{Romano2020OpenBTE:Equation}.

The simulation domain is discretized by a Delaunay mesh bounded by a rectangle of sizes $L_x=L$ along $\mathbf{\hat{x}}$ and $L_y=L$ along $\mathbf{\hat{y}}$; we assume an applied difference of temperature $\Delta T$ = 1 K along the \textit{x}-axis (hence we have a hot and cold side). Once Eq.~\ref{new_bte} is solved, we define the effective ``effective'' 2D thermal conductivity as
\begin{equation}
    \kappa^{\mathrm{2D}}_{\mathrm{eff}}=-\frac{L_x}{L_y}\frac{1}{\Delta T}\int_0^{L_y} \mathbf{J}^{\mathrm{2D}}\cdot\mathbf{\hat{n}}\, dy,
\end{equation}
where the integral runs over either the hot or cold side. As in our case $V_n$ has the units of area, the physical units of $\mathbf{J}^{\mathrm{2D}}$ and $\kappa^{\mathrm{2D}}_{\mathrm{eff}}$ are different from the 3D case, and are W$K^{-1}$ and Wm$^{-1}$, respectively. This argument is consistent with the ``sheet thermal conductance'' introduced by Wu et al~\cite{Wu2016HowThickness}. In macroscopic 2D materials, i.e. with no size effects, these two quantities are related by the 2D Fourier's law $\mathbf{J}^{\mathrm{2D}}=-\kappa^{\mathrm{2D}}\nabla T$. The traditional effective thermal conductivity can be obtained by $\kappa_{\mathrm{eff}} = \kappa_{\mathrm{eff}}^{\mathrm{2D}}/h $, where $h$ is an effective thickness. In the following, all the results are presented in terms of $\kappa_{\mathrm{eff}}$.  

Bulk related data, such as $A_{\mu\nu}$, group velocities, frequencies, and heat capacities are computed by means of the density functional theory implemented in QUANTUM ESPRESSO~\cite{Giannozzi2009fp}, and taken from~\cite{chiloyan2017micro}. The bulk thermal conductivity ($\kappa^{xx}$) is computed within RTA and using the full collision operator (we will refer to this case simply as ``Full''), giving $\approx$ 546 Wm$^{-1}$k$^{-1}$ and 3888 Wm$^{-1}$K$^{-1}$, respectively, with an effective thickness $h = 3.35 \AA$. In absence of size effects, the left hand side of Eq.~\ref{new_bte} is constant and the bulk thermal conductivity is obtained by multiplying both sides of Eq.~\ref{new_bte} by $\mathbf{W}^{\sim1}$, where $\sim1$ is the Moore-Penrose inverse, giving $- \sum_\nu \mathbf{W}^{\sim1}_{\mu \nu \beta} S_\nu^{\beta} \partial_\beta T_\nu = T_\mu $. The total heat flux, therefore, becomes $J^{\alpha}= -\sum_{\mu\nu} \kappa_{\mu\nu}^{\alpha\beta}\partial_\beta T = -  \kappa^{\mathrm{2D}}_{\alpha\beta} \partial_\beta T $, where 
\begin{equation}\label{kappa}
   \kappa_{\mu \nu}^{\alpha \beta} = S_\mu^{\alpha}\mathbf{W}^{\sim1}_{\mu \nu} S_\nu^{\beta},
\end{equation}
is named the ``cross-mode'' thermal conductivity; such a quantity can be interpreted as the heat dissipated on mode $\mu$ when a gradient on the mode temperature $T_\nu$ is applied. The bulk thermal conductivity is then given compactly by $\kappa_{\alpha\beta}^{\mathrm{2D}} = \braket{S^{\alpha}|W^{\sim1}|S^{\beta}}$. In passing, we note that within RTA, in Eq.~\ref{final}  $ W^{\mathrm{od}}_{\mu \nu} = 0 $ \cite{Ziman2001}; in this case $\kappa_{\mu\nu}^{\alpha\beta}  = (V_n N) S_\mu^{\alpha} \left(\tau_\mu / C_\mu \right) S_\mu^{\beta}\delta_{\mu\nu}= (V_n N)^{-1}C_\mu v_\mu^\alpha v_\mu^\beta \tau_\mu \delta_{\mu \nu} $ recovering kinetic theory.

\section{Graphene nanoribbons}

We first apply our formalism to graphene nanoribbons with infinite lengths and varying width $L$. As shown in the inset of Fig.~\ref{fig0a}, the simulation domain is a square with side $L$ with both periodic boundary conditions and a temperature gradient imposed along the \textit{x}-axis. This condition is ensured by setting $T_\mu^{L} = T_\mu^{R} + \Delta T $, where $T_\mu^{L(R)}$ is the mode temperature at the left (right) side of the simulation domain. Note that the heat flux is periodic since $\sum_\mu \mathbf{\mathbf{S}}_\mu \Delta T = 0$. Along the $\textit{y}$-axis, we assume rough surface, i.e. where phonons lose memory upon scattering and are bounced back diffusively. In practice, we set phonons leaving the pores' boundary, $T_B$, to a weighted average of all the incoming flux~\cite{romano2019diffusive}, $T_B = \sum_\mu \alpha_\mu T_\mu/\sum_\mu \alpha_\mu$, where $ \alpha_\mu =  C_\mu \mathbf{v}_\mu \cdot \mathbf{\hat{n}} \mathcal{H}(\mathbf{v}_\mu \cdot \mathbf{\hat{n}} )$. The term $\mathcal{H}$ is the Heaviside function and $\mathbf{\hat{n}}$ is the normal to the surface. This boundary condition is also applied along any internal surface, such as the pores' boundary.

\begin{figure*}

\subfloat[\label{fig1a}]
{\includegraphics[width=0.48\textwidth]{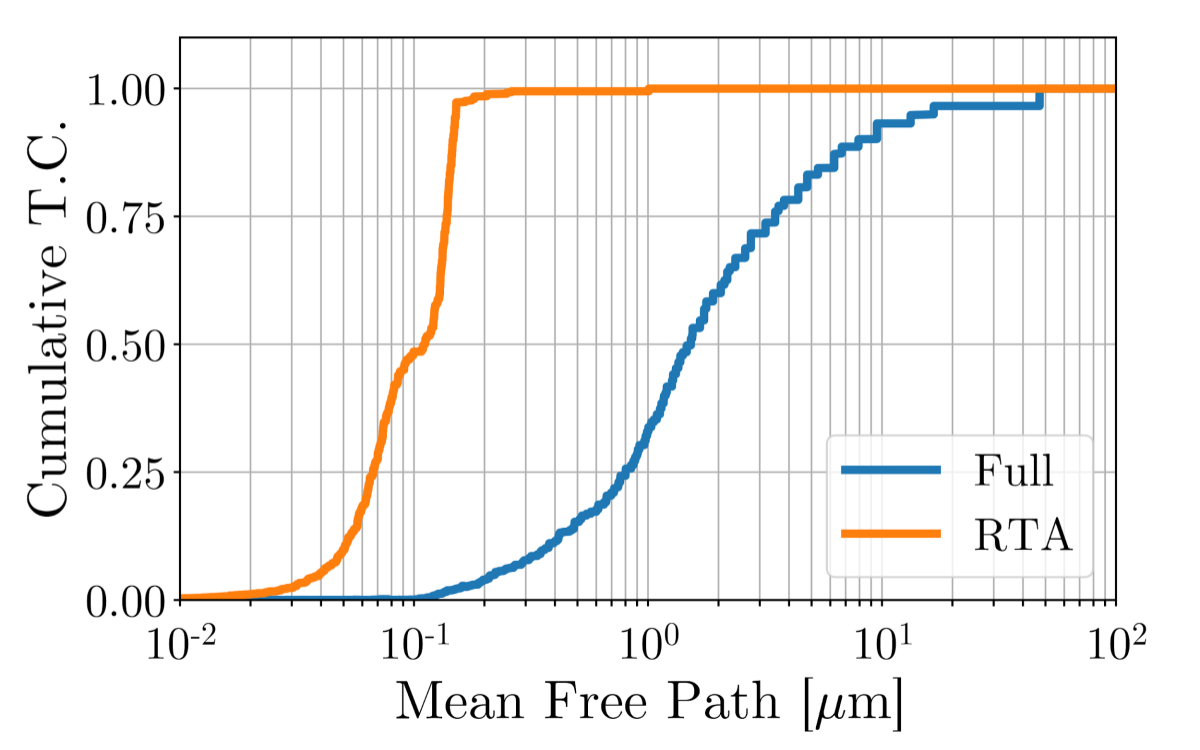}}
\hfill
\subfloat[\label{fig1b}]
{\includegraphics[width=0.48\textwidth]{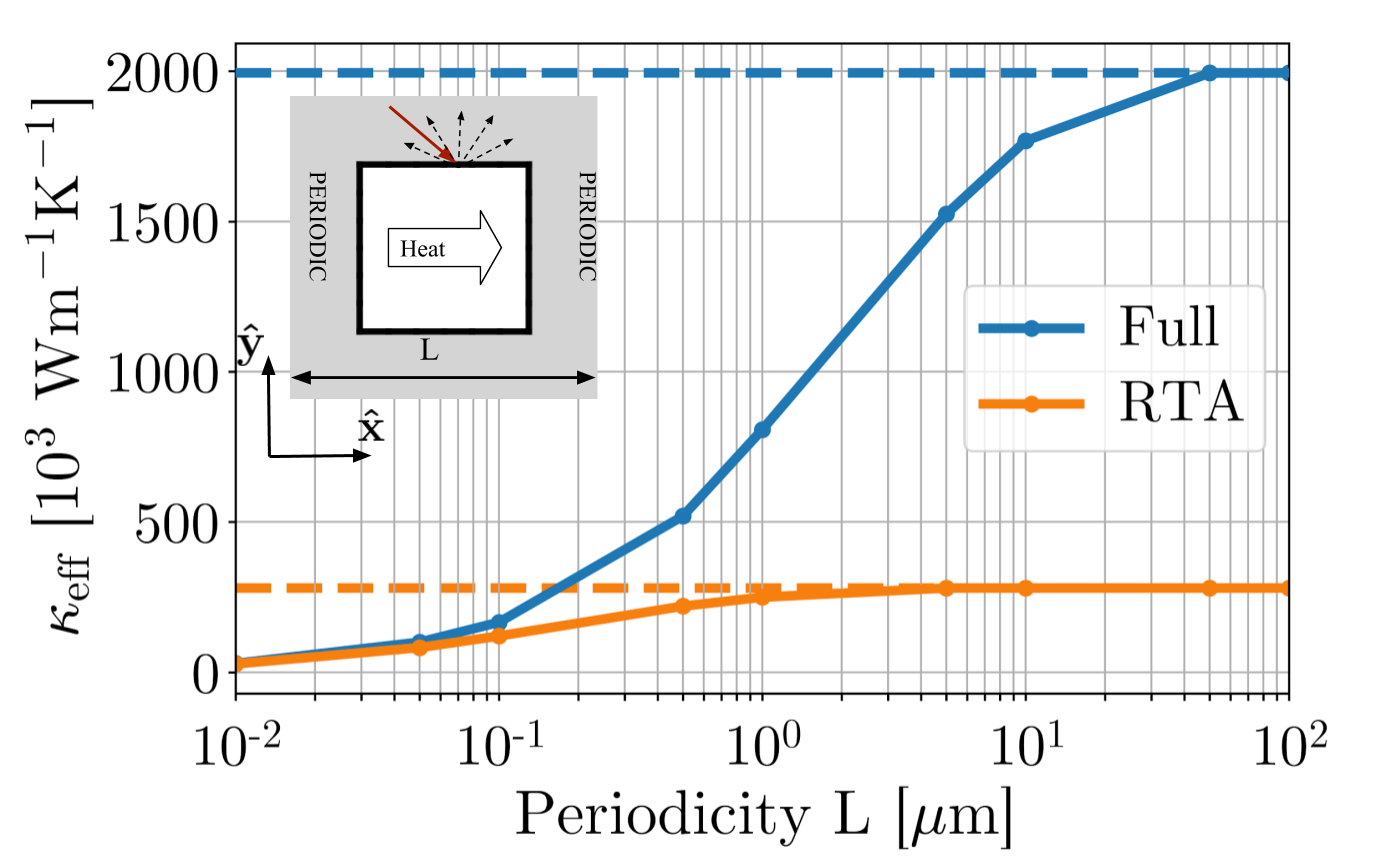}}
\quad
\subfloat[\label{fig1c}]
{\includegraphics[width=0.48\textwidth]{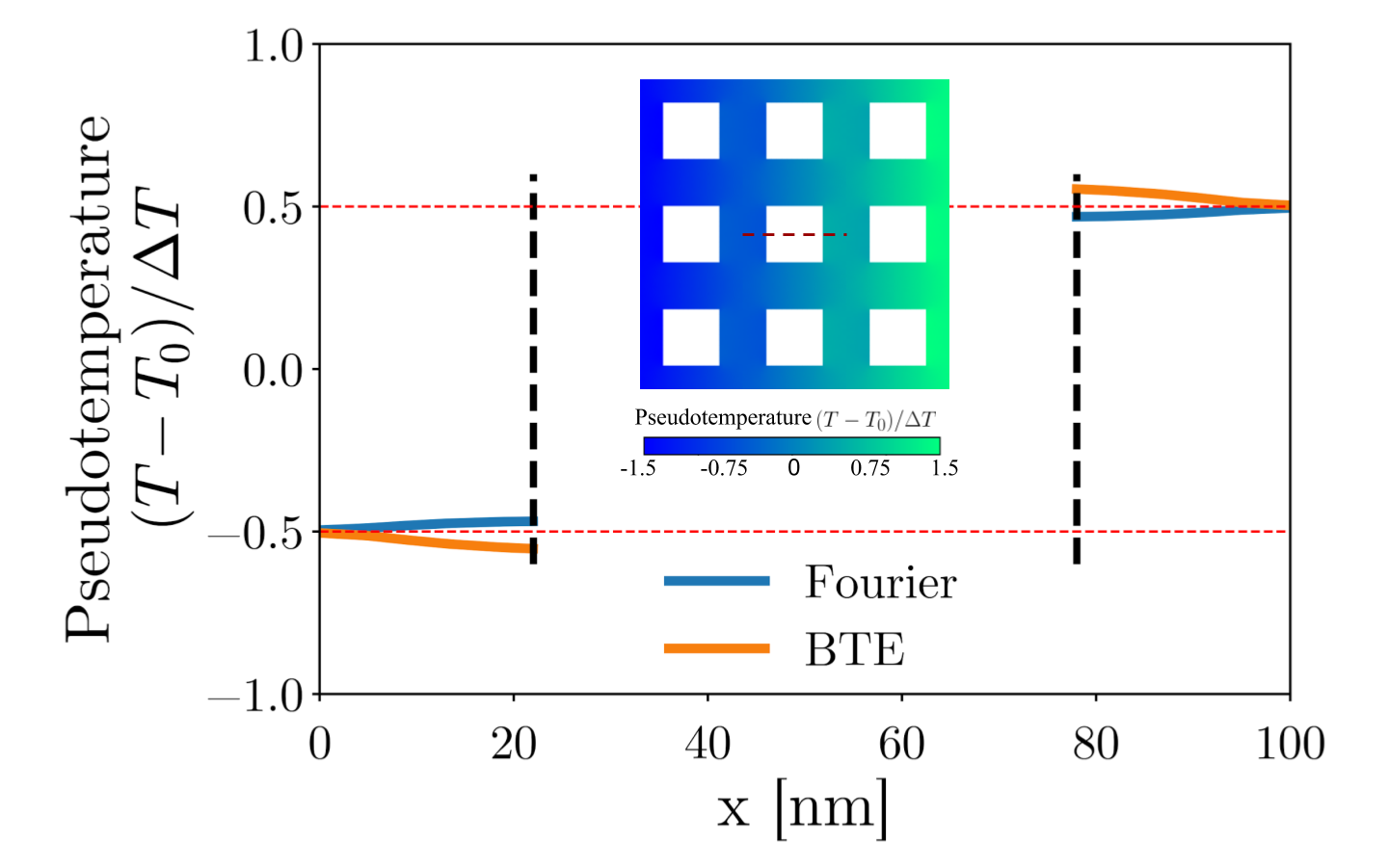}}
\hfill
\subfloat[\label{fig1d}]
{\includegraphics[width=0.48\textwidth]{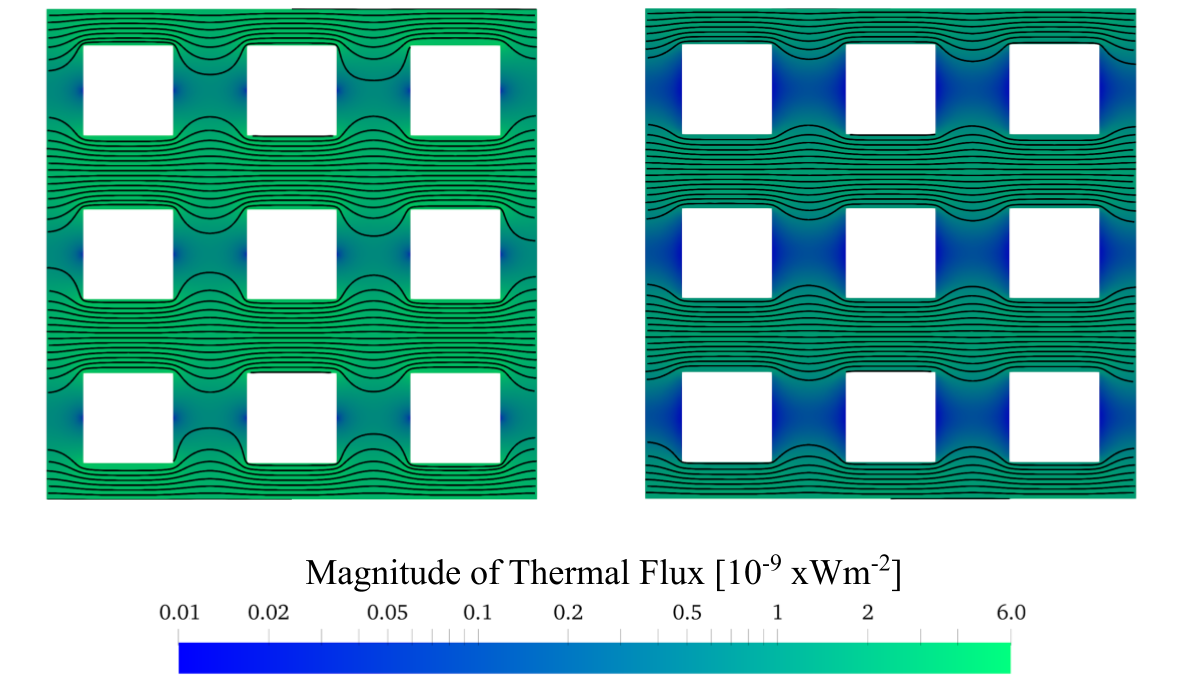}}
\caption{a) Cumulative thermal conductivity for both RTA and Full cases. b) Effective thermal conductivity of porous graphene with aligned pores for different periodicity $L$. As depicted in the inset, we consider a periodic unit-cell comprising a single square pore. Diffuse boundary conditions are applied along the boundary of the walls. The porosity is $\phi$ = 0.3. c) A cut of the temperature $T$ across the \textit{x} for L = 500 nm. In the inset the temperature map is shown. d) Magnitude of thermal flux (log scale) for the Fourier (left) and Full (right) cases.}
\end{figure*}

Fig.~\ref{fig0a} shows $\kappa_{\mathrm{eff}}$ for different nanoribbons' widths. For both RTA and Full cases we observe the ballistic-to-diffusive crossover~\cite{Bae2013BallisticRibbons}, i.e. the transition from a width-dependent $\kappa^{\mathrm{eff}}$ to a regime where heat transport depends only on macroscopic geometric effects. For nanoribbons, this limit is the bulk thermal conductivity. In the Full case, the crossover occurs at a threshold, $L_{\mathrm{diff}}$, that is much larger than that for the RTA case. In fact, from Eq.~\ref{kappa} we note that when the full collision operator is used, heat transport in bulk is dictated by the ``generalized'' phonon MFPs $|\mathbf{G}| = |\sum_\nu \mathbf{W}^{\sim1}_{\mu\nu}\mathbf{S}_\nu| $~\cite{fugallo2014thermal,li2014shengbte}. As shown in Fig.~\ref{fig1a}, the cumulative bulk thermal conductivity distribution versus $|\mathbf{G}|$ spans several micrometers whereas the maximum $|\mathbf{F}|$ of heat-carrying phonons is only a few hundred nanometers. When the off-diagonal terms are related to Normal scattering, i.e. momentum-conserving collisions, this effects can be regarded as phonon hydrodinamic transport~\cite{Cepellotti2017BoltzmannEffect,lee2015hydrodynamic}. The interaction with boundaries for widths much larger than phonon MFPs reflects also in the heat flux profile, which decreases close to the walls even for L = 10 $\mu$m (see Fig.~\ref{fig0b}). This trend is in agreement with Montecarlo simulations~\cite{landon2014deviational} and consistent with the concepts of friction length applied to molybdenum disulfide nanoribbons~\cite{cepellotti2015phonon,Cepellotti2017BoltzmannEffect}.

\section{Porous Graphene}
We investigate thermal transport in porous materials with square pores arranged in a regular lattice. In absence of size effects, $\kappa_{\mathrm{eff}}$ in these systems is solely determined by the porosity, via the Eucken-Garnett formula~\cite{nan1997effective} $\kappa_{\mathrm{eff}} \approx \kappa_{xx} \frac{1-\varphi}{1-\varphi}$ = 2093 Wm$^{-1}$k$^{-1}$ (294 Wm$^{-1}$k$^{-1}$ for RTA). Within RTA, size effects occur when the pore-pore distance becomes comparable to the MFPs~\cite{romano2016directional,song2004thermal}; however, in light of the above discussion, boundary scattering may impact thermal transport even for larger feature sizes, should the off-diagonal terms of the collision operator be significant. 

As shown in Fig.~\ref{fig1a}, thermal transport computed with the Full method reaches the diffusive limit around $L$ = 10 $\mu$m, which is consistent with the distribution of the generalized MFPs. For periodicities as large as 1 $\mu$m, which corresponds roughly to pore-pore distance of 435 nm, we obtain roughly a two-fold reduction with respect to the diffusive limit. Another effect of nondiffusive behaviour is given by the pseudotemperature map, shown in Fig.~\ref{fig1c}. In fact, a cut along the applied temperature difference (which is positive) shows a positive gradient in the distance between the pores, clearly violating Fourier's law. Lastly, the thermal flux, shown in the bottom inset of Fig.~\ref{fig1d}, is pronounced in the space between the pores along the temperature gradient, a signature of ballistic transport. These results, which are consistent with previous literature based on the RTA~\cite{Duncan2019ThermalGratings,hao2009frequency}, can be exploited at much larger scales than the phonon MFPs, thanks to the significant off-diagonal term of the scattering operator~\cite{fugallo2014thermal,cepellotti2015phonon}; as a consequence, low-thermal conductivity materials and high thermal routing capabilities can be achieved via mesoscale patterning. 

\begin{figure*}
\subfloat[\label{fig3a}]
{\includegraphics[width=0.48\textwidth]{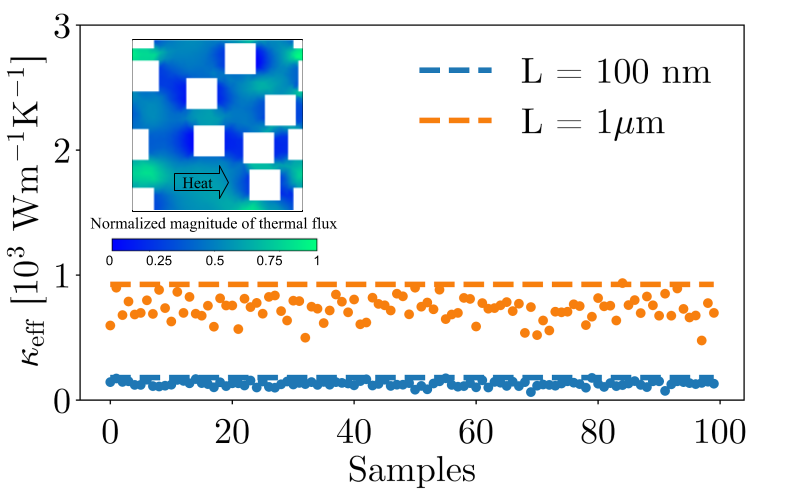}}
\hfill
\subfloat[\label{fig3b}]
{\includegraphics[width=0.48\textwidth]{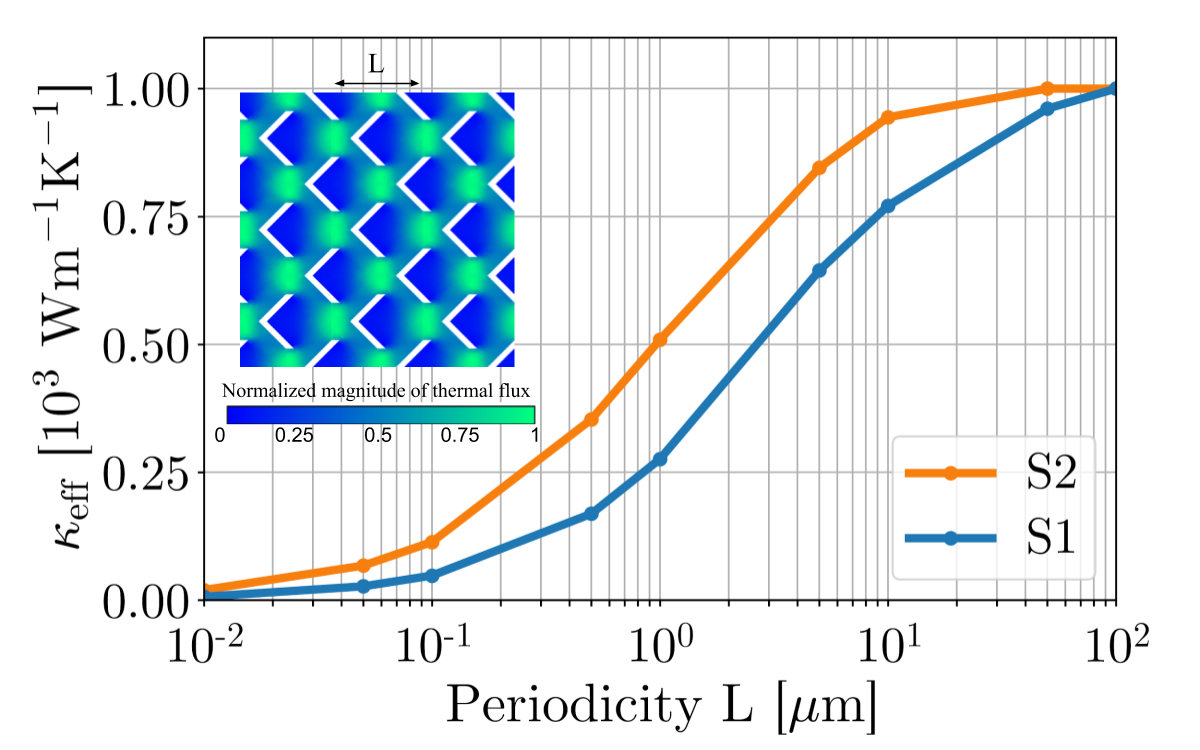}}
\caption{a) Effective thermal conductivity for randomnly chosen configurations for L = 100 nm and 1$\mu$m. The inset shows the magnitude of the thermal flux for the case with the lowest $\kappa_{\mathrm{eff}}$. b) The normalized effective thermal conductivity $\tilde{\kappa} = \kappa_{\mathrm{eff}}/ \kappa_{D}$ (with $\kappa_{D}$ being the diffusive limit) of S1 and S2 structures, with different periodicity $L$. The chosen porosity is $\phi$ = 0.12. The inset shows the magnitude of thermal flux for L = 100 nm.}
\end{figure*}
Similarly to~\cite{romano2016tuning}, we assess whether non aligned configurations may lead to further thermal conductivity reduction. To this end, we consider disordered porous geometries with porosity $ \varphi=0.3 $. Each configuration comprises nine pores randomnly distributed in \textit{x}- and \textit{y} axes. We choose two periodicities, L = 300 nm and L = 3 $\mu$m, for which the pores size are identical to the aligned cases with L = 100 nm, and L = 1 $\mu$m, respectively. We run 100 simulations for each $L$, finding that disordered configurations may lead to a stronger phonon suppression with respect to the aligned counterpart, as shown in Fig.~\ref{fig1c}. This result is mainly due to the reduction in the \textit{view factor}, i.e. the possibility of phonons to travel between the hot and cold contact with no scattering with the boundaries. As depicted in the inset of Fig.~\ref{fig3a}, heat flux follows a tortuous path along the applied temperature gradient, as opposed to the aligned configuration whereas phonons mostly travel through direct paths. Interestingly, for the case with L = 3 $\mu$m, there is a larger variation in $\kappa_{\mathrm{eff}}$; this effect is due to the interplay between the wider distribution of feature sizes and the generalized MFP distribution.



In this last part, we report an example of engineering patterned graphene, where we aim at zeroing the view-factor and ensuring a minimal distance between pores. This requirement rules out solutions with low mechanical stability and vanishing electrical conductivity, an appealing set of properties for thermoelectric applications. We identify the configuration with staggered ``less than'' signs (S1) as a candidate for this task (see inset of Fig.~\ref{fig3b}). The chosen porosity is $\varphi$ = 0.12. For comparison, we also examine the aligned configuration with square pores (S2) and same porosity. As we want to focus on size effects, we consider the quantity $\tilde{\kappa} = \kappa_{\mathrm{eff}}/ \kappa_{D}$m where $\kappa_{D}$ is the diffusive thermal conductivity. From Fig.~\ref{fig3b}, we see that $\tilde{\kappa}$ of the S1 configuration is consistently smaller than that of the S2 structure, for all $L$; notably, a roughly two-fold difference is obtained for $L$ = 1$\mu$m. Furthermore, from the magnitude of the thermal flux, shown in the inset of Fig.~\ref{fig3b}, we note essentially no flux around the right side of the pore, a signature of phonons shadowing induced by the nanostructure~\cite{Duncan2019ThermalGratings}. This result corroborates the use of convex shapes in minimizing thermal transport in nanostructured materials~\cite{gluchko2019reduction}. 

We point out that in graphene the electronic thermal conductivity is a significant fraction to the total thermal conductivity, amounting to 10 $\%$ at room temperature~\cite{Kim2016TheGraphene}. For this reason, topology optimization of 2D materials for heat manipulation should include electronic size effects. While such a combined treatment warrants further research, it is beyond the scope of this work. 

In this paper we have introduced a formalism that computes heat transport in arbitrarily patterned 2D materials, solving the linearized phonon BTE beyond the RTA. We first applied the method to graphene nanoribbons, highlighting the role of the generalized MFPs in dictating the ballistic-to-diffusive crossover. Then, we moved to porous graphene, obtaining remarkable thermal transport suppression for mesoscale patterning. Finally, we identify a promising structure with low porosity and low thermal conductivity. Our method opens up possibilities for parameter-free design of 2D systems for thermal routing and energy harvesting applications. 
\\
The code will be made available in the next release of OpenBTE~\cite{Romano2020OpenBTE:Equation}. 

\begin{acknowledgements}
The author would like to thank Samuel Huberman for helpful discussions and for providing the DFT input data.

\end{acknowledgements}

\section*{APPENDIX A: The scattering operator} \label{ss1}

The collision operator $\Omega$ can be written as 
\begin{equation}
    \Omega_{\mu\mu'} = A_{\mu \mu'}\left[\bar{n}_{\mu'}(\bar{n}_{\mu'}+1) \right]^{-1},
\end{equation}
where $A$ is a symmetric and positive semidefinite matrix~\cite{fugallo2014thermal}, given by

\begin{eqnarray}
    A_{\mu\mu'} =&  \left[\sum_{lk} \left( P_{\mu k\rightarrow l } + \frac{1}{2}P_{lk\rightarrow \mu} \right) +\sum_k P^{\mathrm{isot}}_{\mu,k}\right]  \delta_{\mu\mu'} - \nonumber \\
    & -\sum_{l} \left[ P_{\mu l\rightarrow \mu'} -  P_{\mu\mu'\rightarrow l} +  P_{\mu' l \rightarrow \mu} \right] + P^{\mathrm{isot}}_{\mu,\mu'};
\end{eqnarray}
the term $P^{\mathrm{isot}}$ is the isotope scattering and $P_{\mu \mu' l\rightarrow \mu''}$ are scattering rates for the coalescent event~\cite{fugallo2014thermal}. The latter is given by

\begin{eqnarray}\label{co}
    P_{\mu\mu'\rightarrow \mu''} & = \frac{2 \pi}{N\hbar^2} \left| V^{(3)}_{\mu,\mu',-\mu''} \right|^2 \bar{n}_\mu \bar{n}_{\mu'}\left(\bar{n}_\mu'' + 1 \right)  \times\nonumber \\ 
                                   \times & \Delta_{\mu,\mu',-\mu''}\delta(\hbar \omega_\mu + \hbar \omega_{\mu'} - \hbar \omega_{\mu''}),
\end{eqnarray}
where $\Delta_{\mu,\mu',-\mu''}$ is one when $\mathbf{q}_\mu + \mathbf{q}_{\mu'} - \mathbf{q}_{\mu''}$ is a reciprocal lattice vector and zero otherwise. Lastly, the term $V^{(3)}$ is the third derivative of the energy with respect atomic positions. For details, see~\cite{fugallo2014thermal}.

\section*{APPENDIX B: Energy Conservation} \label{ss2}

As seen in the main text, the scattering operator needs to satisfy the following conditions 
\begin{equation}\label{cond}
\sum_\mu W_{\mu\nu} = \sum_\nu W_{\mu\nu} = 0
\end{equation}
In practice, however, Eq.~\ref{cond} is not strictly satisfied since the delta functions used in Eq.~\ref{co} are replaced by Gaussians~\cite{landon2014deviational}. To ensure strict energy conservation, hence, the non-conserving scattering matrix, denoted by $\tilde{W}_{\mu\nu}$ has to be modified such that $W_{\mu\nu} = \tilde{W}_{\mu\nu}+\beta_{\mu\nu}$ satisfies Eq.~\ref{cond}. In our case, the lack of energy conservation, estimated by $\sum_\nu \left|\sum_{\mu} W_{\mu\nu}\right|/\sum_{\nu\mu} \left|W_{\mu\nu}\right|$~\cite{Landon2014ARedacted}, is about 6.5 $\%$. The matrix $\beta_{\mu\nu}$ is computed using the method of the Lagrange multipliers~\cite{Landon2014ARedacted}. Specifically, we minimize the function
\begin{eqnarray}
    F = \sum_{\mu\nu}\beta^2_{\mu\nu}+& \sum_\nu \lambda_\nu^r \left(\sum_\mu \beta_{\mu\nu}+\Delta_\nu \right)+ \\ \nonumber
    +& \sum_\mu \lambda_\mu^c \left(\sum_\nu \beta_{\mu\nu}+\Delta_\mu \right),
\end{eqnarray}
where $\Delta_\nu = \sum_\mu \tilde{W}_{\mu \nu} = \sum_\mu \tilde{W}_{\nu \mu} $ (where we used the symmetry of W), and  $\lambda^c_\nu$  ($\lambda^r_\mu$) are the Lagrange multipliers enforcing zero sum for columns (rows). The stationary solution of $F$ is found by solving the following system of equations
\begin{eqnarray}
     \frac{\partial F}{\partial \beta_{m n}} = 2\beta_{m n} + \lambda^r_n + \lambda^c_m = 0 \nonumber \\
     \frac{\partial F}{\partial \lambda_n^r} = \left(\sum_m \beta_{m n}+\Delta_n \right) = 0 \nonumber \\
     \frac{\partial F}{\partial \lambda_m^c} = \left(\sum_n \beta_{m n}+\Delta_m \right) = 0,
\end{eqnarray}
which gives $\beta_{mn} = -\left(\lambda^r_n + \lambda^c_m \right)/2$, $\sum_m \beta_{m n} = -\Delta_n$, and $ \sum_n \beta_{m n} = -\Delta_m$. Combining these solutions, we have $N  \lambda^c_n +  \sum_m \lambda^r_m   = -2\Delta_n $ and  $N \lambda^c_m +  \sum_n \lambda^r_n  = -2\Delta_m $. The Lagrange multipliers are obtained by solving the linear system

\begin{equation}
    \begin{pmatrix}
    \mathbf{I} & \mathbf{G}  \\
    \mathbf{G} & \mathbf{I}
    \end{pmatrix}
     \begin{pmatrix}
     \mathbf{\lambda}^r  \\
     \mathbf{\lambda}^c
    \end{pmatrix}= -\frac{2}{N}
     \begin{pmatrix}
     \mathbf{\Delta}  \\
     \mathbf{\Delta}
    \end{pmatrix},
\end{equation}
where $G_{mn} = 1/N$ and $\mathbf{I}=\delta_{mn}$. We stress that an energy conserving scattering matrix is paramount to the algorithm detailed in the main text. Nevertheless, after applying the constrain, we note little change in the bulk thermal conductivity, i.e. $\braket{S^{\alpha}|W^{\sim1}|S^{\beta}}\approx \braket{S^{\alpha}|\tilde{W}^{-1}|S^{\beta}} $. 

\bibliography{references} 

\end{document}